\documentclass[twocolumn,showpacs,preprintnumbers,amsmath,amssymb]{revtex4}
\usepackage{bm}% bold math
\usepackage{color,graphicx}

\begin{document}

%\preprint{APS/123-QED}

\title{%Relativistic properties of a spin entanglement of two massive Dirac particles using a mean spin operator\\
Mean spin entanglement of two massive Dirac particles under Lorentz transformations}

\author{Taeseung Choi}
\email{tschoi@swu.ac.kr}
\affiliation{ Barom Liberal Arts College, Seoul Women's University, Seoul 139-774, Korea } %\textbackslash\textbackslash
\author{Jin Hur and Jaewan Kim}
\affiliation{School of Computational Sciences, Korea Institute for Advanced Study, Seoul 130-012, Korea }
\date{\today}% It is always \today, today,
             %  but any date may be explicitly specified \\

\begin{abstract}
 We have studied the relativistic effects on the mean spin entanglement of two massive Dirac particles
      using the simultaneous eigen-spinors of the Foldy-Woutheysen mean spin operator and the Dirac Hamiltonian.
      We have obtained the transformation matrix from the spinor with specific momentum to the spinor with
      a transformed momentum under an arbitrary Lorentz transformation.
      Using the transformation matrix we have shown the consistent
      monotonic behavior between the concurrence and the maximum
      value of Bell parameter in Bell inequality of transformed
      spin states.

%The mean spin operator commutes with the Dirac Hamiltonian and becomes the conventional spin operator
%of the Pauli Hamiltonian in non-relativistic limit.

\end{abstract}

\pacs{ }% PACS, the Physics and Astronomy
\maketitle
%\maketitle

\section{Introduction}
Quantum entanglement is a very important physical resource in
quantum computing\cite{Nielson}. A separated pair of entangled
quantum system can be used to transfer quantum information.
 The measurement on one system of an entangled pair determines the outcome of the measurement on the other system instantaneously. Entanglement has a non-local nature in its essence.
 The special theory of relativity, however, tells us that there is no agreement on simultaneity
 for distant events among observers in inertial reference frames moving relative to each other.
 There are no causal relations between two spatially separate events.
 In this aspect relativity is local in its nature.
 Hence it is an intriguing question how the entanglement of a pair changes among
 observers in relative motions.

The spin state of elementary spin $1/2$ particles, such as
electrons, are good candidates for qubits. Massive spin $1/2$
particles, however, have momentum degrees of freedom other than
the spin degrees of freedom. Moreover it has been shown the
relativistic transformation entangles the spin and momentum
degrees of freedom and the spin entropy determined by the reduced
density matrix for the spin is not covariant under Lorentz
transformation \cite{Peres}.
 %It has been shown that quantum entropy is not a Lorentz-covariant concept and how special
% relativity imposes severe constrictions on the transfer of information between distant systems .
 Then the spin entanglements would also depend on the motion of observers and it is interesting how the spin entanglements
 between massive spin $1/2$ particles would be changed under relativistic transformations.
 Several authors have investigated the effect of the entanglement
 under relativistic transformations \cite{Czachor, Alsing, Remb}.

 Gingrich and Adamai \cite{Gingrich}
 concluded that the spin entanglement of a pair of particles is not invariant under relativistic
 motion, although the whole entanglement of spin and momentum degrees of freedom is Lorentz invariant.
 They have founded that the special pairs of spin which are initially separable can be maximally entangled
 in the special reference frame. They also noted that this property could lead
 to simplified state preparation and purification protocols.
 Later Czachor {\it et. al.} have pointed out the definition of reduced density matrix with
 traced-out momenta in Ref. \cite{Gingrich} is not justified.
  The results of later works have not reached the
 same conclusion depending on whether only the change of states are considered \cite{Terashima} or
 what kinds of relativistic spin observables are used \cite{Ahn, Lee, Kim}.
The main reason for the discrepancy is the lack of clear
understanding for the proper spin operator for massive Dirac
particles.

In this letter, we will clarify the problems associated with the
relativistic effects on quantum spin entanglement. The discrepancy
in the previous works will be cured by considering a Dirac
equation and
 spin observables consistently.
 Dirac equation is a successful theory to reconcile quantum mechanics with special relativity.
 Therefore we will use the Dirac theory for a massive spin 1/2 particle, so-called
 a massive Dirac particle,
 to obtain a consistent covariant approach for the relativistic effect on the spin
 entanglement of a pair of Dirac particles.
 We use `mean spin angular momentum' operators, defined by Foldy and Woutheysen\cite{Foldy},
 to describe a good quantum spin observables we will call these
 operators as mean spin operators.
 The mean spin operators commute with Dirac Hamiltonian unlike the spin operator in the conventional representation
 and since the time evolution of the state of Dirac particle is generated by Dirac Hamiltonian,
 the mean spin operators are constants of motion and become good quantum observables.
 The spin operator in the proper non-relativistic Pauli theory is
 this mean spin operator which describes the measured spin of the
 particle.

A free massive spin $1/2$ particle, {\it i.e.} a massive Dirac particle, is described by the time-dependent Dirac equation
 \cite{Dirac},
\begin{eqnarray}
i \frac{\partial }{\partial t}\Psi({\bf x},t)= \mathcal{H}_D \Psi({\bf x},t) = (\beta m + \mbox{\boldmath{$\alpha$}} \cdot {\bf P})
\Psi({\bf x},t),
\end{eqnarray}
where $\mbox{\boldmath{$\alpha$}}$ and $\beta$ are $4\times 4$
Dirac matrices in standard Dirac representation, ${\bf P}$ is a
momentum operator and a natural unit defined by $c=\hbar=1$ is
used. Here we are interested in the change of entanglement by
relativistic motion of observers so that it is enough to consider
the free Dirac particles without additional interactions.

The eigenvalue equation of the Dirac Hamiltonian $\mathcal{H}_D$ is as follows
\begin{eqnarray}
(\beta m + \mbox{\boldmath{$\alpha$}} \cdot {\bf P})\psi^\pm({\bf p},\lambda) = \pm E  \psi^\pm({\bf p},\lambda),
\end{eqnarray}
where  $E=\sqrt{m^2 + {\bf p}^2}$ and the superscript $\pm$
represents positive-energy and negative-energy solutions.
      The state $\psi^\pm({\bf p},\lambda)$ can also be understood as the Lorentz transformed state from the state
  $\psi^\pm( {\bf 0},\lambda)$ in the rest frame of the particle such as
  \begin{eqnarray}
  \label{Eq:DiracSpinor}
  \psi^\pm( {\bf p},\lambda) = \sqrt{\frac{m}{E}}S(L(p))\psi^\pm({\bf 0},\lambda),
  \end{eqnarray}
  %\textcolor{red}{Check!! upper equation in Sakurai}
  where $p^\mu = \sum_\nu L^\mu_{\phantom{\mu} \nu}k^\nu$.
  $L^\mu_{\phantom{\mu} \nu}$ is a standard Lorentz transformation depending on $p^\mu$ and
  $S(L(p))$ is a 4-spinor representation for $L(p)$.
  We use the following normalization
  \begin{eqnarray}
  &&(\psi^+( {\bf p'}, \lambda'), \psi^+( {\bf p},\lambda)) =
  \delta_{\lambda' \lambda} \delta^3({\bf p}'-{\bf p}) \\ \nonumber
  &&(\psi^-( {\bf p'}, \lambda'), \psi^-( {\bf p},\lambda)) =
  \delta_{\lambda' \lambda} \delta^3({\bf p}'-{\bf p}),
  \end{eqnarray}
  where $(,)$ is a scalar product in Hilbert space and the scalar products between other states are zero.
  Note that $p^0 \delta^3({\bf p}'-{\bf p})$ is a Lorentz invariant delta function.

Here the label $\lambda$ represents other degrees of freedom than momentum degrees of freedom.
It must be related to the spin of the particle, however
 at this stage, the meaning of this label is not clear since the conventional $4\times 4  $ spin operators
 $
 \frac{1}{2}\mbox{\boldmath{$\sigma$}} = \frac{1}{4i}\mbox{\boldmath{$\alpha$}} \times \mbox{\boldmath{$\alpha$}}
 $
 does not commute with the Hamiltonian $\mathcal{H}_D$.
 The proper spin operator must commute with the Hamiltonian which governs the dynamics of
 the particle and also satisfy the SU(2) algebra.
The operator $\frac{1}{2}\mbox{\boldmath{$\sigma$}}$ commutes with the
Hamiltonian $\beta m$
 of the rest frame of the particle and satisfies SU(2) algebra, hence the proper spin operator must be reduced to
 $\frac{1}{2}\mbox{\boldmath{$\sigma$}}$ in the rest frame of the particle.
 Foldy and Woutheysen(FW) has defined the mean spin operator by the use of the ``canonical'' transformation
 such as \cite{Foldy}
\begin{eqnarray}
%\mathbf{\Sigma} = e^{-i S} \mbox{\boldmath{$\sigma$}}  e^{i S},
\frac{1}{2}\mathbf{\Sigma} = U^\dagger_{FW}({\bf P}) \frac{1}{2}\mbox{\boldmath{$\sigma$}} U_{FW}({\bf P}),
\end{eqnarray}
where $U_{FW}({\bf p})$ is the unitary operator as
\begin{eqnarray}
U_{FW}({\bf P})=\frac{m+ \beta \mbox{\boldmath{$\alpha$}}\cdot {\bf P}+ E}{\sqrt{2E(E+m)}}.
%&& S = -i \beta  \left(\mbox{\boldmath{$\alpha$}} \cdot \hat{\mathbf{p}}\right) w, \quad \hat{\mathbf{p}}= \frac{\mathbf{p}}{|\mathbf{p}|} \\
%&& w= \frac{1}{2} \tan ^{-1}\left(\frac{|\mathbf{p}|}{m}\right).
\end{eqnarray}
%The transformation $e^{i S}$ is unitary since $S$ is Hermitian.
The mean spin operators satisfy SU(2) algebra and commute with the Dirac Hamiltonian, i.e.,
%\begin{eqnarray}
$
 \left[ \frac{1}{2}\mathbf{\Sigma}, \mathcal{H}_D \right] = 0.
 $
%\end{eqnarray}
Therefore it is possible to find simultaneous eigenstates of
$\mathcal{H}_D$ and $\Sigma_3$, such as
\begin{eqnarray} \nonumber
&& \mathcal{H}_D \psi^\pm_{FW}( {\bf p},\lambda)=\pm E \psi^\pm_{FW}( {\bf p},\lambda), \quad \left(E=\sqrt{m^2+\mathbf{p}^2}\right),
\\
&& \Sigma _3 \psi^\pm_{FW}( {\bf p},\lambda)=\lambda  \psi^\pm_{FW}( {\bf p},\lambda), \quad (\lambda =\pm 1).
\end{eqnarray}
It is now clear that the variable $\lambda$ in $\psi^\pm_{FW}( {\bf p},\lambda) $
represents the $z$ component of the mean spin operator $\mathbf{\Sigma}$.
The expectation value of the mean spin operator describes the average spin of the particle
which is really measured. This mean spin operator is the same as the spin operator
in the non-relativistic Pauli Hamiltonian when it is represented in the representation (so-called
Foldy-Woutheysen representation) where the Dirac Hamiltonian has the diagonal form.
Note that for positive-energy states the conventional Dirac spinor $\psi^+({\bf p},\lambda)$
is the same as the Foldy-Woutheysen spinor $\psi^+_{FW}( {\bf p},\lambda)$.
%and the mean spin operator $\frac{1}{2}\mbox{\boldmath{$\Sigma$}}$ reduces to $\frac{1}{2}\mbox{\boldmath{$\sigma$}}$
%in the rest frame.
This implies the mean spin for a particle with positive energy in its rest frame does not change
under the Lorentz transformation $L(p)$.
%\textcolor{red}{It seems strange since an observer sees a particle with definite momentum
%not the particle's history of Lorentz transformation. SO if there was no momentum entanglement at first
%there will be no entanglement changes!!!
%Therefore the mean spin entanglement of a pair of partcles
%in the rest frame preserves under only one standard Lorentz boost.
%The rotation of an observer does not affect the entanglement of two particles.

The general situation appears when the observer to see a particle with momentum $p^\mu$
moves relativistically such that the momentum of the particle becomes $\sum_\nu\Lambda^\mu_{\phantom{\mu} \nu} p^\nu$,
where $\Lambda$ is an arbitrary orthochronous Lorentz transformation. The spinor state $\psi^\pm_{FW}( {\bf p},\lambda)$ described by the observer in old reference frame will be transformed as
$S(\Lambda) \psi^\pm_{FW}( {\bf p},\lambda)$ to the observer in the new reference frame,
where $S(\Lambda)$ is the spinor representation for $\Lambda$.

It is nontrivial to represent $S(\Lambda) \psi^\pm_{FW}( {\bf p},\lambda)$ with eigenstates
$\psi^\pm_{FW}( \Lambda{\bf p},\lambda)$ in the new reference frame
since the two successive Lorentz transformations $S(\Lambda)$ and $S(L(p))$ does not become one
Lorentz transformation in general.
Here $\Lambda {\bf p}$ is a spatial component for the transformed 4-momentum.
A FW-representation will be used to represent the transformed state explicitly.
In the FW-representation, the transformed Dirac Hamiltonian and the mean spin operator take simple forms
\begin{eqnarray}
\mathcal{H}_D' &=& U_{FW}({\bf P}) \mathcal{H}_D U_{FW}^\dagger({\bf P}) = \beta (m^2+\mathbf{P}^2), \\
\frac{1}{2} \mathbf{\Sigma}' &=& U_{FW}({\bf P}) \frac{1}{2} \mathbf{\Sigma} U_{FW}^\dagger({\bf P}) = \frac{1}{2}\mbox{\boldmath{$\sigma$}}.
\end{eqnarray}
Note that both operators are block diagonal. This fact makes that the eigenstates in FW-representation can be written as the product form
\begin{eqnarray}
\psi'^\pm (\mathbf{p},\lambda) = | {\bf p} \rangle \otimes |\lambda\rangle^\pm.
\end{eqnarray}
Now the eigenstate in the original representation can be obtained as
%\begin{eqnarray}
$\psi^\pm_{FW} (\mathbf{p},\lambda) = U_{FW}^\dagger({\bf P}) \psi'^\pm (\mathbf{p},\lambda)$.
%\end{eqnarray}
%\color{black}
It is convenient to use the new notations for the spin state $|\lambda\rangle^\pm$ in FW-representation such as
$|+\rangle^{+}= \left( \begin{array}{cccc} 1 &0 &0  & 0 \end{array} \right)^T\equiv |0\rangle$, $
|-\rangle^{+}\equiv |1\rangle$, $|+\rangle^{-}\equiv |2\rangle$, and $
|-\rangle^{-}\equiv |3\rangle.$

 We define a transformation matrix $\mathcal{T}^{(\Lambda,p)}$ which represents the
 %\textcolor{red}{index is ${\bf p}$ or $p$???}
transformation of a FW-spinor state under an arbitrary Lorentz transformation $\Lambda$:
\begin{eqnarray}
\label{Eq:TRMatrix}
S(\Lambda)\psi_{FW}( {\bf p}, \mu)=\sum _{\nu} \mathcal{T}^{(\Lambda,{\bf p})}_{\nu \mu } \psi_{FW}(\Lambda {\bf p},\nu),
\label{Udef}
\end{eqnarray}
where $\mu, \nu=0,1,2,3.$
This relation is described in the FW-representation as follows
\begin{eqnarray*}
S(\Lambda )U^\dagger_{FW}({\bf P}) |{\bf p} \rangle \otimes |\mu \rangle=\sum _{\nu} \mathcal{T}^{(\Lambda,{\bf p})}_{\nu \mu }
U^\dagger_{FW}({\bf P}) |\Lambda {\bf p} \rangle \otimes |\nu \rangle.
\end{eqnarray*}
%where $\Lambda {\bf p}$ represents the spacial momentum 3-vector transformed from ${\bf p}$ by $\Lambda$ .
Therefore the transformation matrix is obtained as
\begin{eqnarray}
\mathcal{T}^{(\Lambda,{\bf p})}= \langle \Lambda {\bf p}| U_{FW}({\bf P}) S(\Lambda) U^\dagger_{FW}({\bf P})| {\bf p}\rangle.
\end{eqnarray}

To find an explicit form of $\mathcal{T}^{(\Lambda,{\bf p})}$, we consider simple Lorentz boost $\Lambda _{\xi }$
in $x_3$ direction with rapidity $\xi$. It is clear that a rotation of the space does not have any effect on the entanglement. %Since we are interested in the change of entanglement after Lorentz transformation,
Therefore we consider only a Lorentz boost without loss of generality.
Writing the space part of momentum vector in spherical coordinate $
%\begin{eqnarray}
p = \{E,p \sin{ \theta}  \cos{ \phi },p \sin {\theta } \sin {\phi
},p \cos {\theta }\}, $
%\end{eqnarray}
where $\theta$ is a polar angle from the positive $x_3$-axis and $\phi$ is the azimuthal angle in the $x_1 x_2$ plane from
the $x_1$-axis.
We get
\begin{eqnarray}
\mathcal{T}^{\left(\Lambda _{\xi },p\right)}=\sqrt{\frac{E'}{E}} \left(
\begin{array}{cccc}
 A & B e^{-i \phi } & C & D e^{-i \phi } \\
 -B e^{i \phi } & A & D e^{i \phi } & -C \\
 0 & 0 & \tilde{A} & \tilde{B} e^{-i \phi } \\
 0 & 0 & -\tilde{B} e^{i \phi } & \tilde{A}
\end{array}
\right), \label{Ufinal}
\end{eqnarray}
where
\begin{eqnarray}
A&=& \sqrt{\frac{m+E}{m+E'}} \left[\cosh \left(\frac{\xi }{2}\right)+\frac{p \cos (\theta )}{m+E} \sinh \left(\frac{\xi }{2}\right)\right], \\
B&=& \frac{p \sin (\theta )}{\sqrt{(m+E) \left(m+E'\right)}} \sinh \left(\frac{\xi }{2}\right), \\
C&=& \frac{\sinh \left(\frac{\xi }{2}\right)}{E'\sqrt{(E+m) \left(E'+m\right)}} \left[\cosh ^2\left(\frac{\xi }{2}\right) \left\{(m+E)^2 \right.\right. \nonumber\\
&& \qquad \left.\left.-p^2 \cos (2 \phi )\right\}+m p \cos (\phi ) \sinh (\xi )\right], \\
D&=& -\frac{p \sin (\phi) \sinh (\xi )}{E' \sqrt{(E+m) \left(E'+m\right)}} \left[E \sinh \left(\frac{\xi }{2}\right) \right. \nonumber\\
&& \qquad \left. +p \cos (\phi ) \cosh \left(\frac{\xi }{2}\right)\right], \\
\tilde{A} &=& \frac{E}{E'} A, \qquad \tilde{B}= \frac{E}{E'} B, \\
E'&=& p \sinh (\xi ) \cos (\theta )+E \cosh (\xi ).
\end{eqnarray}

Since the Dirac Hamiltonian in the FW-representation is in diagonal form,
positive energy subspace is spanned by the upper two components of the state and
negative energy subspace is spanned by the lower two components of the state.
The sign of energy is invariant under Lorentz transformation so that
we do not have to consider negative energy state for a particle with positive mass $m$.
Here we will study the entanglement of particles with positive energies and
all considerations can be given in the two-dimensional positive energy subspace.
Therefore the concurrence is a good measure of the spin entanglement for two massive
positive energy Dirac particles.
In the positive energy subspace the transformation matrix becomes $2\times 2$ matrix
\begin{eqnarray}
\mathcal{T}_2^{\left(\Lambda _{\xi },p\right)}=\sqrt{\frac{E'}{E}} \left(
\begin{array}{cc}
 A & B e^{-i \phi } \\
 -B e^{i \phi } & A \end{array}
 \right). \label{TM2}
\end{eqnarray}

A general positive energy state in consideration can be written as
\begin{eqnarray}
\Psi= \int d^3 {\bf p} \sum_\lambda a_{{\bf p},\lambda} \psi^{(+)}_{FW}({\bf p},\lambda).
\end{eqnarray}
Where the normalization condition is satisfied as
$(\Psi,\Psi)= \int d^3 {\bf p}\sum_\lambda {\mid {a_{{\bf p},\lambda}} \mid}^2.$
The density matrix $\rho$ is defined by $\Psi \Psi^\dagger$.
Then the $\sigma \sigma'$-component of the reduced spin density matrix
$\rho^r$ is obtained by
\begin{eqnarray}
\rho^r_{\sigma, \sigma'} = \int d^3{\bf p}(\psi^{(+)}_{FW}({\bf p},\sigma), \rho\psi^{(+)}_{FW}({\bf p},\sigma')).
\end{eqnarray}
Note that this reduced density matrix is $2\times 2$ matrix as expected for the positive energy state.
%\begin{widetext}
%\begin{multline}
%\rho^r_{\sigma, \sigma'} =\int d^3{\bf p}''\left( \psi^{(+)}({\bf p},\sigma),
%\int d^3 {\bf p}' \sum_\lambda a_{{\bf p},\lambda} \psi^{(+)}({\bf p},\lambda)
% \int d^3 {\bf p} \sum_\lambda' a^*_{{\bf p},\lambda'} \psi^{(+)\dagger}({\bf p},\lambda')
% \psi^{(+)}({\bf p},\sigma')\right)
% \end{multline}
%\end{widetext}
%The spin entanglement given by the concurrence is then obtained from Wootters formula \cite{Wootters}
The spin entanglement of this reduced state is obtained from the
concurrence of Wootters formula
%\begin{eqnarray}
$
C = \max\{\lambda_1-\lambda_2-\lambda_3-\lambda_4,0\}
$
%\end{eqnarray}
\cite{Wootters, Note},
where $\lambda_i$'s are the square roots of the eigenvalues of the matrix $\rho\tilde{\rho}$ with
%\begin{eqnarray}
$
\tilde{\rho} = (\sigma_y \otimes \sigma_y) \rho^* (\sigma_y \otimes \sigma_y).
$
%\end{eqnarray}

It is interesting to study the effect of Lorentz boost on the
state whose reduced spin state is a Bell state initially,
%Consider a state
%\begin{eqnarray}
%\Psi _{AB}=\frac{1}{2} \left(\left
%|{\bf p}_1,-{\bf p}_1\right\rangle +\left|{\bf p}_2,-{\bf p}_2\right\rangle \right) (|1,1\rangle +|-1,-1\rangle ), \label{initialstate}
%\end{eqnarray}
\begin{eqnarray}
&& \Psi_{AB} = \frac{1}{2} \left\{ \Psi({\bf p}_1,-{\bf p}_1;1,1)+ \Psi({\bf p}_1,-{\bf p}_1;-1,-1) \right. \nonumber\\
&& \qquad + \left. \Psi({\bf p}_2,-{\bf p}_2;1,1) + \Psi({\bf p}_2,-{\bf p}_2;-1,-1) \right\},
\end{eqnarray}
where \\
$\Psi({\bf p}_A,-{\bf p}_B;\lambda_A,\lambda_B) =
\psi_{FW}^{(+)}({\bf p}_A,\lambda_A) \otimes \psi_{FW}^{(+)}({\bf
p}_B,\lambda_B)$
 and ${\bf p}_1$, ${\bf p}_2$ are in the $x_1
x_2$-plane with the same magnitude $p$ and spin variables
represents the eigenvalues of mean spin operators. The angle
between ${\bf p}_1$ and ${\bf p}_2$ is $\phi$.
%The state $\Psi _{AB}$ represents pair of
%particles with opposite momentum. When one particle has momentum
%${\bf p}_1$(${\bf p}_2$), the other particle has momentum $-{\bf
%p}_1$($-{\bf p}_2$).
By writing the state as a density matrix and
tracing over the momentum variable, %we get a reduced density
%matrix $\rho^r_{AB}$. If we trace over momentum for (\ref{initialstate}),
we get a maximally entangled spin state ($C=1$). For an observer
moving to the positive $x_3$ axis with rapidity $\xi$, the state
changes according to the transformation matrix (\ref{TM2}) and,
therefore, the concurrence would change also.

%A result is shown in Fig. \ref{fig1} for $\phi=\pi/2$.
%\begin{figure}
%\includegraphics[scale=1.3]{fig1.eps}\\
%\caption{Concurrence $C$ as a function of rapidity $\xi$. We chose $m=1$ and $\phi=\frac{\pi}{2}$ for both plot.}\label{fig1}
%\end{figure}
%This figure shows the concurrence decreases as $p$ and $\xi$ increases.
%Fig. \ref{fig1} shows the general behavior of the concurrence depending on the magnitude of momentum of particles
%$p$, rapidity $\xi$, and the angle $\phi$.
%When the magnitude of particle momentum and
% rapidity increases there appears zero concurrence for a certain value of the angle $\phi$.
 %This angle goes to $\pi/2$ and a concurrence curve will have
 %reflection symmetry with the axes of symmetry at $\phi=\pi/2$ in the limit of $p, \xi \to\infty$.
%The concurrence reaches to a different limit value depending on the magnitude of the initial momentum
%as the rapidity $\xi$ increases.
 Taking limit $p\to\infty$ and $\xi\to\infty$, the concurrence yields the simple form
\begin{eqnarray}
\label{Eq:Conc}
C = |\cos(\phi )|.
\end{eqnarray}
In this limit the final reduced spin state has all possible
concurrence depending on the angle $\phi$. When ${\bf p}_1$ and
${\bf p}_2$ are perpendicular, the spin entanglement vanishes
entirely.

 For completeness we consider the Bell inequality \cite{Bell, CSHS} as a parallel approach.
 The Bell inequality violation is given only for states with non-zero concurrence,
 since the entanglement is needed to violate the Bell inequality. Moreover for two qubit systems
 the amount of violation of Bell inequality can be an entanglement measure to some extent \cite{Munro}.
 Therefore we expect the consistent violation of Bell inequality is obtained with the concurrence.
 %in Eq. (\ref{Eq:Conc}) for the Lorentz-transformed spin state with the limit of $p, \xi \to\infty$.
 The previous results \cite{Terashima, Kim, Lee}, however, shows the maximal violation of
 Bell inequality is obtained in case the proper direction of spin measurement is chosen.
 We will show the consistent results are obtained independent on the direction of spin
 measurements different from the previous results. The main reason
 for this is that we consider both the transformation of the state
 and a spin operator correctly.

\begin{figure}
\label{fig2}
\includegraphics[scale=1.3]{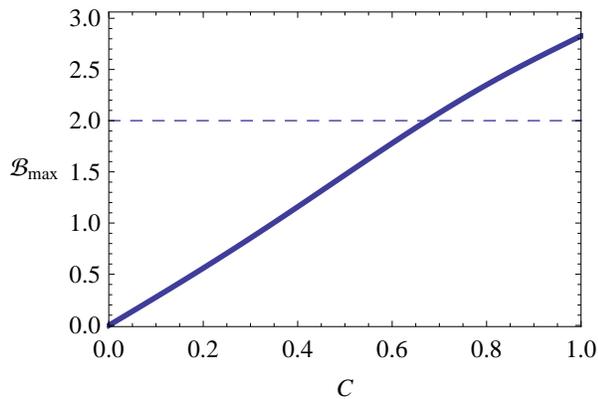}\\
\caption{The maximum value of Bell parameter
$\mathcal{B}_{\text{max}}$ versus the concurrence for the reduced
spin states in the limit of $p, \xi \to\infty$.}
\end{figure}

 The generalized Bell inequality is determined by the following Bell parameter,
\begin{eqnarray}
\mathcal{B} = \mid C({\bf a}_1, {\bf b}_1)+  C({\bf a}_1, {\bf b}_2)+  C({\bf a}_2, {\bf b}_1)
-  C({\bf a}_2, {\bf b}_2)\mid,
\end{eqnarray}
where the spin correlation function is given by
%\begin{eqnarray}
$
 C({\bf a}, {\bf b}) = \mbox{Tr}\{ \left( \frac{1}{2}\mathbf{\Sigma}_A \cdot {\bf a} \right)\otimes
 \left(\frac{1}{2}\mathbf{\Sigma}_B \cdot {\bf b} \right)\rho^r\},
 $
%\end{eqnarray}
where $\mathbf{\Sigma}$ is the mean spin operator and $\bf a$($\bf
b$) is the direction for spin measurement. For a given state the
maximum value of Bell parameter, $\mathcal{B}_{\text{max}}$, is
obtained by adjusting the directions ${\bf a}_1$, ${\bf b}_1$,
${\bf a}_2$, and ${\bf b}_2$ for spin measurement. Fig. \ref{fig2}
shows $\mathcal{B}_{max}$ versus the concurrence in Eq.
(\ref{Eq:Conc}) for the reduced spin state in the limit of $p, \xi
\to\infty$. This shows one-to-one correspondence between two
quantities, specifically, a monotonic behavior. The maximum value
of Bell parameter becomes less than 2 even though the entanglement
remains ($C \neq 0$). This is because the more entanglement is
required to give the same Bell parameter for mixed states
\cite{Munro}. The reduced spin state for $\phi=\pi/2$ in the limit
of $p, \xi \to\infty$ is the mixed state, $\frac{1}{2}
|\Psi_-\rangle \langle \Psi_-| + \frac{1}{2} |\Phi_+\rangle
\langle \Phi_+|$, where $|\Psi_-\rangle$ and $|\Phi_+\rangle$ are
Bell states. The concurrence and the maximum value of Bell
parameter for this state are both zero.

We have studied the relativistic effects on the spin entanglement
of two massive Dirac particles using the simultaneous eigenstate
of the Dirac Hamiltonian and the mean spin operator. In this
formalism the meaning of spin operators as quantum observables is
clear.
%We have clarified the meaning of the label $\lambda$ which
%represents the other degrees of freedom than a momentum in Dirac
%spinor. The $\lambda$ is the eigenvalue of the mean spin operator.
The mean spin operators commute with the Dirac Hamiltonian and
become the conventional spin operators of the Pauli Hamiltonian in
non-relativistic limit \cite{Foldy}. This means the mean spin
operator is the good quantum variable involved in spin
measurement. We have found the transformation matrix which
represents the transformation of a spinor state with given
momentum to spinor states with the transformed momentum under an
arbitrary Lorentz transformation.
%It is shown the concurrence is
%the good measure of spin entanglement for positive energy states
%since the positive-energy subspace is two-dimensional and not
%mixed with negative-energy subspace under Lorentz transformations.
%We have shown that for the initial spin Bell state after tracing
%over the momentum degrees of freedom the best choice for Alice and
%Bob to share the spin entangled pair is to send each particle to
%Alice and Bob in opposite direction.
We have shown the consistent monotonic relation between the
concurrence and Bell inequality of the final reduced spin states
transformed from the special initial states. We expect this
consistent behavior is achieved for other states considering our
mean spin correlations. We have clarified the meaning of trace
over momentum for a 4-spinor state, so
%Our study have clarified how the 4-spinor
%related with real spin measurement transforms under the arbitrary
%Lorentz transformation.
%The resultant concurrence is the same as that in ref. \cite{Gingrich}.
%This coincidence comes from the fact that
%for the positive-energy state the Lorentz transformed state $\psi^+({\bf p},\lambda)$ in Eq. (\ref{Eq:DiracSpinor})
%is the same as the Foldy-Woutheysen eigenspinor $\psi^+_{FW}( {\bf p},\lambda)$.
%This fact, however, does not mean the $\lambda$ in the state $\psi^+({\bf p},\lambda)$ is
%the spin eigenvalue for the conventional spin operator $\frac{1}{2}\sigma_z$.
%\textcolor{red}{Please Check!! which is better position of the above comments. Notice... or just right above??}
our approach makes it possible to study
 the relation between the entanglement of momentum and spin of two Dirac particles
 under the arbitrary Lorentz transformation in clear manner.

%\begin{acknowledgments}
 This work was supported by National Research Foundation of Korea Grant funded by the Korean Government(KRF-2008-331-C00073)
 and by National Research Foundation of Korea Grant funded by the Korean
 Government(2010-0004855).
 We gratefully acknowledge KIAS members for helpful discussions.
%\end{acknowledgments}

\begin{references}
\bibitem{Note} Notice that the concurrence of the states considered in ref. \cite{Gingrich},
is the same as that of our calculation. This coincidence comes
from the fact that for the positive-energy state the Lorentz
transformed state $\psi^+({\bf p},\lambda)$ in Eq.
(\ref{Eq:DiracSpinor}) is the same as the FW eigenspinor
$\psi^+_{FW}( {\bf p},\lambda)$. This fact, however, does not mean
the $\lambda$ in the state $\psi^+({\bf p},\lambda)$ is the spin
eigenvalue for the conventional spin operator
$\frac{1}{2}\sigma_z$. In our approach the entanglement of the
whole system having both spin and momentum degrees of freedom is
not invariant in general.
\end{references}
%\begin{thebibliography}{99}

%\end{thebibliography}

\end{document}